# Methods for Linking Data to Online Resources and Ontologies with Applications to Neurophysiology

Matthew Avaylon, Ryan Ly, Andrew Tritt, Benjamin Dichter, Kristofer E. Bouchard, Christopher J. Mungall, and Oliver Rübel*

**Scientific Data Division, Lawrence Berkeley National Laboratory, Berkeley, CA, USA**
Oliver Rübel, Andrew Tritt, Ryan Ly, Matthew Avaylon, Kristofer Bouchard

**Environmental Genomics and Systems Biology, Lawrence Berkeley National Laboratory, Berkeley, CA, USA**
Christopher J. Mungall

**CatalystNeuro**
Benjamin Dichter

*corresponding author, oruebel@lbl.gov, ORCID 0000-0001-9902-1984

# Abstract

Across many domains, large swaths of digital assets are being stored across distributed data repositories, e.g., the DANDI Archive [8]. The distribution and diversity of these repositories impede researchers from formally defining terminology within experiments, integrating information across datasets, and easily querying, reusing, and analyzing data that follow the FAIR principles [15]. As such, it has become increasingly important to have a standardized method to attach contextual metadata to datasets. Neuroscience is an exemplary use case of this issue due to the complex multimodal nature of experiments. Here, we present the HDMF External Resources Data (*HERD*) standard and related tools, enabling researchers to annotate new and existing datasets by mapping external references to the data without requiring modification of the original dataset. We integrated *HERD* closely with Neurodata Without Borders (NWB) [2], a widely used data standard for sharing and storing neurophysiology data. By integrating with NWB, our tools provide neuroscientists with the capability to more easily create and manage neurophysiology data in compliance with controlled sets of terms, enhancing rigor and accuracy of data and facilitating data reuse.

# 1. Introduction

To successfully understand neuroscience data and especially the work from other labs, context is everything. There has been a steady shift in the neuroscience field towards the sharing and the reuse of data for analysis. While data standards (e.g., NWB and BIDS [19]), in conjunction with data archives (e.g., DANDI or OpenNeuro [18]), have significantly advanced FAIR sharing of neuroscience data, the lack of rigorous definition of metadata terminologies and a mechanism to uniquely identify metadata entities remain key challenges for data reuse. This creates the need to relate terms from datasets and experiment attributes to community standard definitions.

We define the term *external resources* as web-accessible resources (e.g., ontologies, brain atlases, gene and model organism databases, data archives, or scholarly resources) that uniquely identify terms and assets, providing highly detailed and precise information about specific topics. Integrating data linkages to external resources provides valuable context for future analysis, ensuring FAIR data use, **Figure 1**. For example, to describe the species of a subject, a user may use a broad range of terms to represent the same thing, such as “human” or “homo sapiens.” To avoid ambiguity and enable search and integration of data across files, we need to define the meaning of such terms. Fortunately, the biosciences community has created many well-curated, readily-available ontological resources to address this challenge.

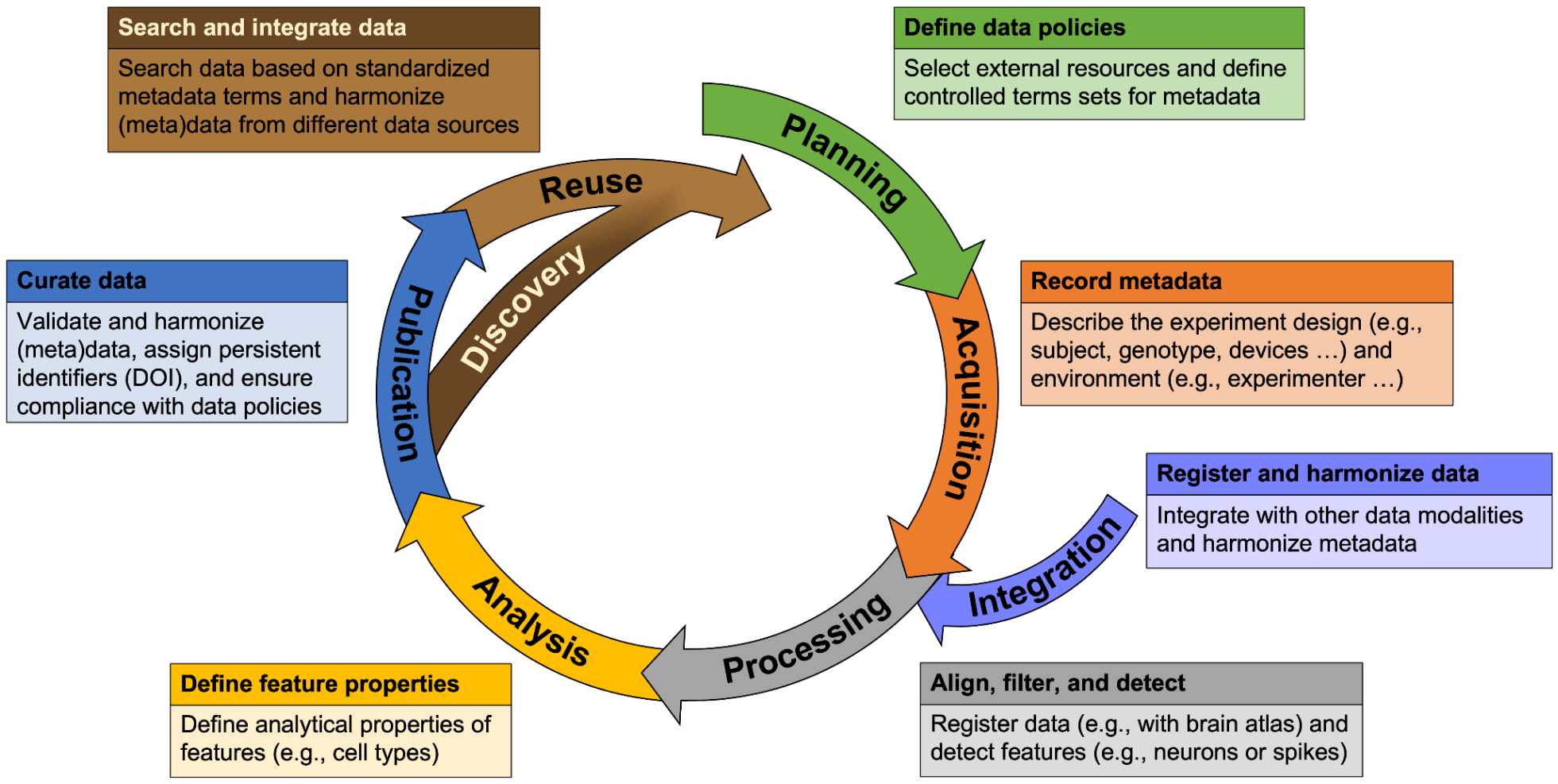

**Figure 1** Illustration of the phases of the neuroscience data lifecycle (arrows) and typical tasks that require external resources.

From the perspective of a data standard, a key challenge is that it is often impossible to *a priori* specify which resources should be used. For example, depending on the species of a subject, we may need to use different brain atlases to standardize brain coordinates. Further compounding the issue, there are often multiple resources available with overlapping scope, and users frequently need to use multiple external resources simultaneously to ensure rigorous data specification. As such, it is often not feasible to require the use of specific external resources in the schema of the data standard. Instead, data producers need to be able to dynamically link (i.e., upon creation of a file) *keys* (e.g., user terms) stored in *objects* (e.g, datasets and attributes within a file) to *entities* (e.g., terms) of online resources.

To address this challenge, we extend the Hierarchical Data Modeling Framework (HDMF) [1] to support creating, managing, and storing metadata references to ontological resources and controlled vocabularies. Though the framework is agnostic to a specific data standard, we focus the implementation of this work around the Neurodata Without Borders (NWB) [2] data standard. NWB supports both HDF5 [7] and Zarr [9] as file format backends for storing complex neurophysiology data. NWB is a community driven data standard and researchers from around the world have taken advantage of the NWB framework to integrate publicly available analysis tools and to extend the scope of NWB via custom neurodata extensions (NDX).

The key novel contributions of this manuscript are:

1. We introduce a common format for defining controlled terminologies based on LinkML (Sec. 2.1.1)
2. We define the novel *HERD* data model for linking data and metadata to external resources (Sec. 2.1.2)
3. We define a spectrum of methods for integrating controlled set of terms with neuroscience data standards to meet the needs of the broad set of use case throughout the data livecycle (**Fig. 1**), ranging from rigorous enforcement of terminologies as part of data standards to dynamic annotation of existing data for harmonization (Sec 2.2). We further describe the implementation of this approach as part of NWB.

4. We demonstrate the application of our tools to a select set of use cases in the context of NWB (Sec. 3).

# 2. Methods

## 2.1 Technological Foundations

### 2.1.1 How to Define Controlled Terminologies

In order to communicate meaning, it is essential to establish and use consistent terminology with rigorous definitions to support accurate interpretation of data. Within HDMF, we created *TermSet* to provide an intuitive and easy-to-use mechanism to create and store a curated, reusable set of terms and define the meaning of terms by linking them to rigorous definitions in ontologies and identities via persistent identifiers. A *TermSet* defines an enumeration of permissible terms, where each defined term is a relation from a user defined key to an ontology specific ID and corresponding URL for the term in the ontology. We use LinkML [4] to support rigorous definition of the term within a *TermSet*, specifically taking advantage of LinkML Enumerations to define said controlled vocabulary in a LinkML YAML schema. We use LinkML, because their modeling language and tools are well integrated with ontologies, and it presently supports integration with a wide range of data modeling technologies (including JSON-Schema, ShEx, RDF, OWL, GraphQL, and SQL DDL).

To create a *TermSet*, users start with defining a YAML schema [6] that holds all the permissible values and corresponding sources. This schema follows the structure defined by the LinkML enumeration model, in which the sources are linked with a standardized prefix, **Figure 2**. Users can manually create complete enumerations of permissible terms. However, this approach is not scalable and hard to maintain when the collection of valid terms is large. Dynamic enumerations and *SchemaSheets* in LinkML provide user-friendly tools to simplify the creation of term sets via user-friendly spreadsheets and through automatic extraction of terms from ontologies. For a complete example of a *TermSet* YAML schema, refer to **Appendix A.1**.

```yaml
# Set prefix for OBO Cell Ontology
prefixes:
   CL: http://purl.obolibrary.org/obo/CL_
```

```yaml
enums:
  NeuronTypeEnum:
    reachable_from:
      source_ontology: obo:cl
      source_nodes:
        - CL:0000540
      include_self: false
      relationship_types:
        - rdfs:subClassOf
```

**Figure 2** (*left*) The syntax for setting the prefix for sources in the schema. (*right*) The syntax for setting up dynamic enumerations.

Dynamic enumerations solves the issue of manually setting permissible terms for large vocabularies by allowing users to populate the terms according to a supported ontological source generatively. There are hundreds of ontologies in which users are able to choose from a variety of sources: i) BioPortal, storing over 700 biomedical ontologies, ii) the OBO Foundry, which contains a broad range of ontologies for the biological sciences, iii) AberOWL, a repository of biological ontologies, and many more. As illustrated in **Figure 2**, we assign the *source_ontology* key in the schema to the source that houses the ontology (in this case the OBO Foundry) as well as the ontology abbreviation (in this case the Cell Ontology). Users can use the entire ontology; however, for both specificity and less memory intensive schemas, the *source_nodes* parameter selects a node in the ontology as a root with all child nodes being pulled for permissible terms. Once the schema is set, the *TermSet* class in HDMF is able to generate a fully expanded vocabulary of the designated nodes, stored as a new schema with populated permissible values. For a complete example, refer to **Appendix A.2**.

Many research labs have ontologies and reference metadata stored locally across spreadsheets. The *TermSet* class is also integrated with *SchemaSheets* [17], a LinkML tool to create and populate a schema from a Google spreadsheet. Using *SchemaSheets*, users can then export the definitions of the schema as one or more tsv files. Using the generated tsv files, our *TermSet class* can then automatically generate the schema by resolving the metadata in the tsv files for required fields, as well as all permissible terms in the enumeration.

### 2.1.2 How to Link Data and Metadata to External Resources

Here we introduce the methodology of *HERD* as a data standard to create and manage linkages between metadata and neurophysiology datasets and attributes stored within the NWB format. Conceptually, one can think of *HERD* as a data model for storing large collections of links to external resources in a flat, denormalized table, in which each row associates a particular key stored in a particular object (i.e., an Attribute or a Dataset in an NWB file) with a particular entity (e.g., a term) of an online resource (e.g., an ontology). However, to ensure normalized data storage without data duplication and to make data queries efficient; *HERD* stores these assets internally in an SQL-style collection of interlinked normalized tables. As illustrated in **Figure 3**, the overall structure of *HERD* resembles that of a simple relational database.

In *HERD*, we need to be able to relate keys, i.e., terms, to data objects in a particular file. Subsequently, those keys need to be linked to entities from an external online resource. In order to accurately keep track of where the data exists, *HERD* stores the unique identifier from a NWB file internally in the *FileTable*, denoted as *files* in **Figure 3**. Recall that *HERD* annotations are for datasets and attributes within NWB objects. The *objects* table stores not only the link between the object and the file it resides in, but also the type of the object and the relative path to the targeted field, i.e., a dataset or attribute. For datasets with a compound data type where each data element consists of multiple components, we can further identify which component with the *field* column. The values from the data or applicable attributes are used as keys, refer to the *keys* table in **Figure 3**, that are then mapped to an entity, consisting of the entity identifier and URI from the resource. The keys are values from fields within objects. We store this association with the *object_keys* table. Similarly, the linkage between the keys and their formal definition, i.e, the entity association, is stored in *entity_keys.*

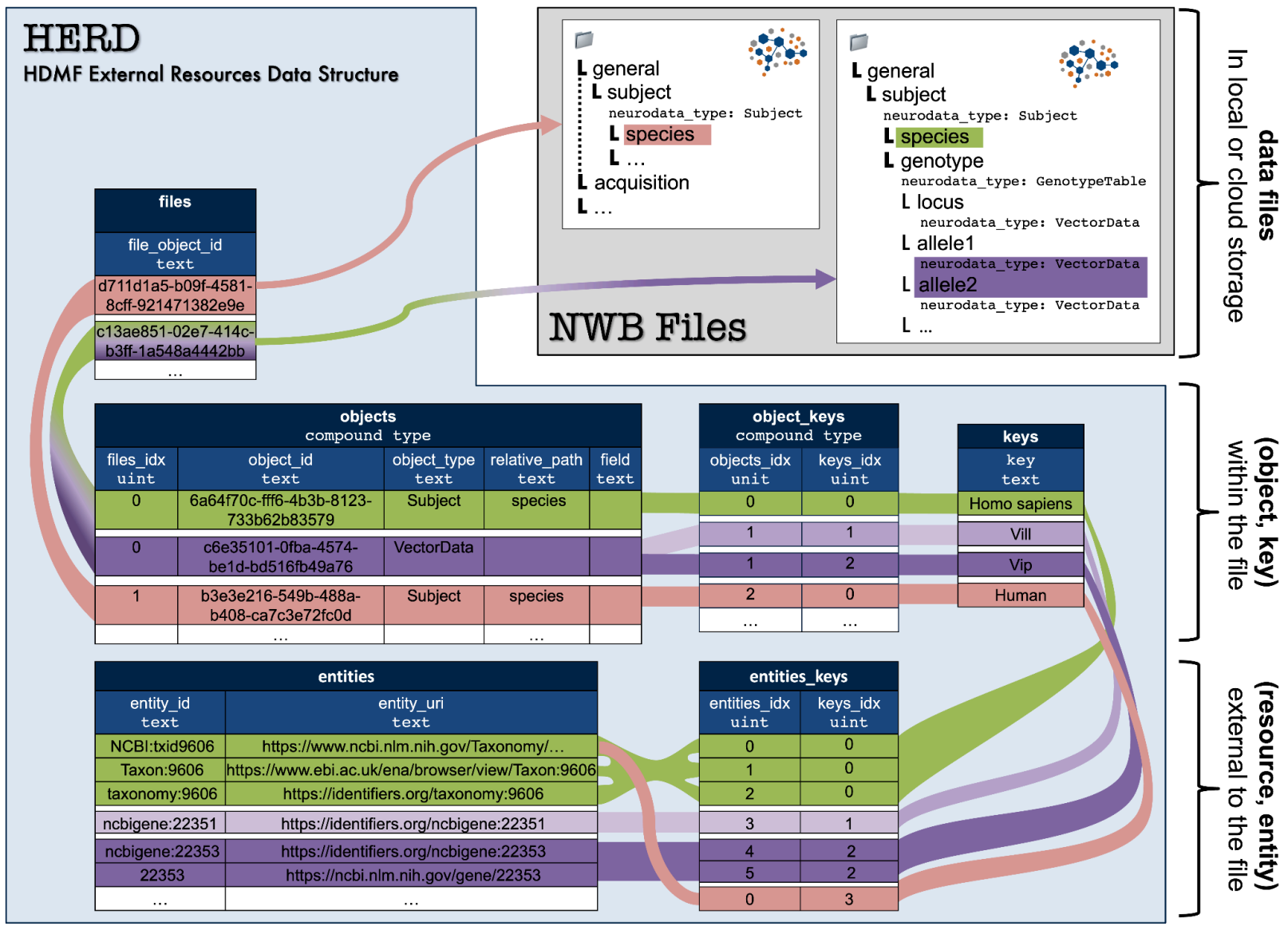

**Figure 3** Overview of the HDMF External Resources Data Structure (*HERD*) for linking data stored in NWB files to external, online resources.

By defining objects, keys, and entities as separate tables and linking the tables via the *object_keys* and *entity_keys* tables enables us to share keys across objects as well entities across multiple keys while ensuring normalized data storage (i.e., without having to duplicate information). This approach supports efficient and easy key-, object-, and entity-based queries—e.g., to identify all objects that share a particular key or all keys that link to the same entity—and harmonization of data. For example, key objects with names Human and Homo sapiens would be linked to the same entity NCBI_TAXON9609 as part of the NCBI Taxonomy [10], i.e., even though the files themself store different keys we can now consistently query the subject by querying based on the unique enity_id. Note, in contrast to keys and entities, objects are unique to a file such that we can store the files_idx directly as part of the *objects* table. The HDMF API efficiently supports common queries across the *HERD* data structure and conversion of query results (and the whole data structure) to flat Pandas dataframes for convenient analysis.

With so much data already available, it is essential that any system that provides meaningful ontological references for neuroscience experiments supports these existing datasets. By having the *HERD* data structure stored externally from NWB files, we avoid having to modify large, existing NWB files when creating these annotations. Currently, *HERD* is written as a zipped collection of tsv files, one tsv for each table of the data structure. By separating the storage of *HERD* files from the actual data, it enables *HERD* to collect metadata linkages for multiple files in a single instance, reducing the number of *HERD* files needed to keep track of external references and facilitating harmonization and integration of data across files.

## 2.2 Methods for integrating controlled term sets with neuroscience data standards

The mechanism for using controlled terms and linking data to external resources depend on a range of factors, e.g,. i) does the user need to annotate an already existing dataset or a new dataset while it is being generated, or ii) is the data for a novel experiment that is still being developed or is the data for a standardized experiment, e.g., as part of a larger science consortium. These sorts of questions affect both: i) the level of rigor in the definition and enforcement of controlled terms as well as ii) the level of automation by which annotations for linking data to external resources can be generated (see **Fig. 4**). More specifically, the types of mechanisms for annotating data that are applicable, depend in practice often on the particular phase in the data lifecycle. The further along we are in the data lifecycle, the more flexible the methods for defining terminology typically need to be. Similarly, the particular mechanisms that are applicable also depend on the maturity and scale of the experiment. Novel, single-lab experiments that are still in development usually require highly flexible definitions of terms whereas large-scale and mature projects require highly automated and rigorous mechanisms for using and enforcing common terminologies. Next, we will discuss the implementation of the four methods shown in **Fig 4** (blue boxes) in the context of a common application use case: 1) manually annotating existing data files (Sec. 2.2.1), 2) defining terms sets to use for particular data objects to annotate new data files (Sec. 2.2.2), 3) pre-configuring term sets to use for types of objects to design experiments (Sec. 2.2.3), and 4) defining term sets to use in data scheme to enhance rigor in the design of data standards (Se. 2.2.4). .

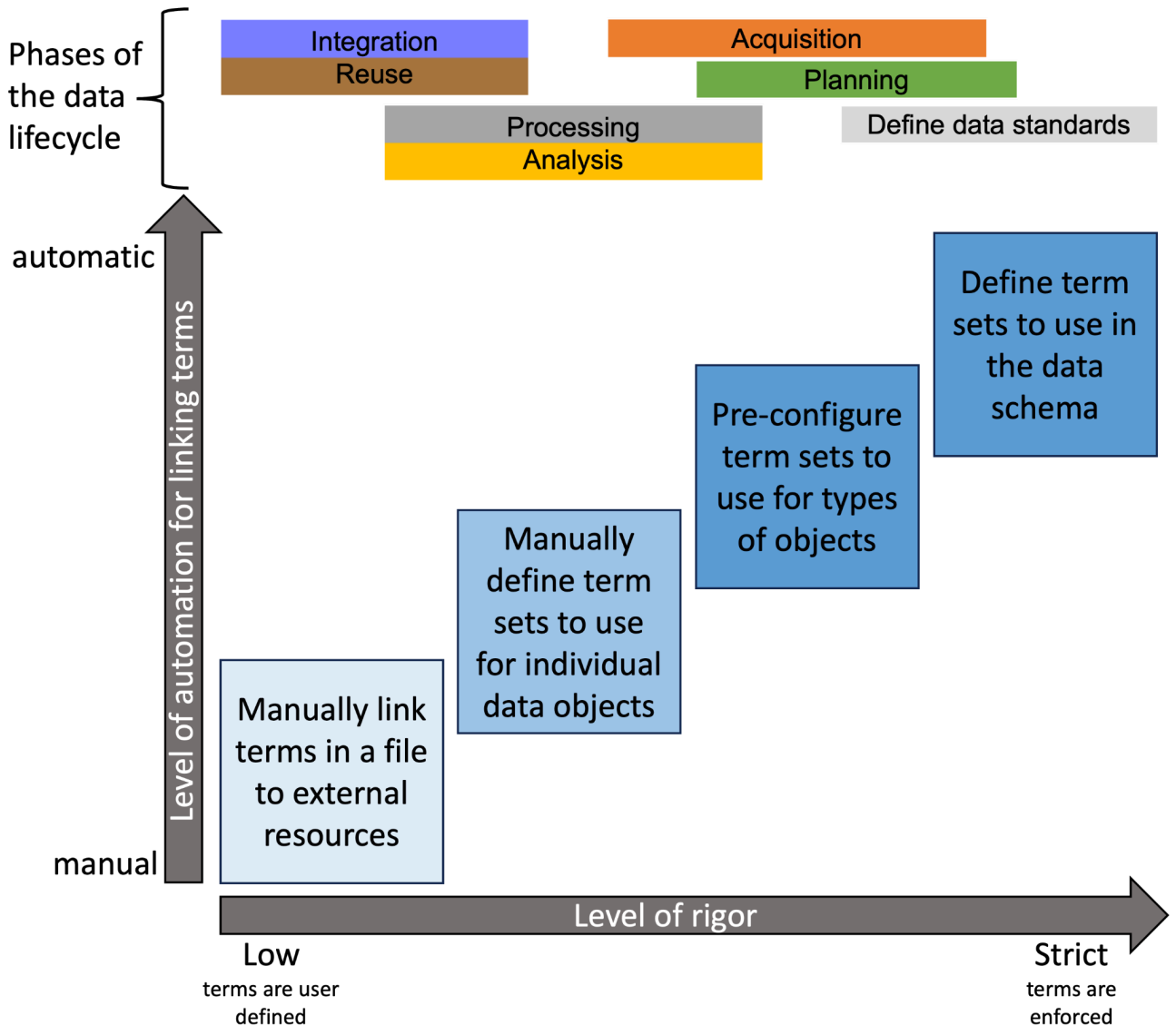

**Figure 4:** Overview of the main methods for linking terms to external resources (y-axis) and enforcing the use of controlled terms as part of the data (x-axis).

### 2.2.1 How to Annotate Existing Data and Attributes

When dealing with data already present in an NWB file, we want to avoid having to rewrite the file itself to add contextual metadata. Users will use the *HERD* class in HDMF directly to manually define links of terms to external resources via the *add_ref* or *add_ref_termset* methods. The *add_ref* method is a low-level function to populate a single “row” in the *HERD* data structure, granting the user greater control. This method internally resolves the rules that define and standardize the *HERD* data structure, allowing users to populate it as if it were a single table. As shown in **Figure 5** (step 1), the user defines: i) the file that contains the object, ii) the object itself, iii) the attribute that is the target for the link, iv) the key that holds the value of the target (this is the value itself from the attribute), and v) the explicit assignments of the *entity_id* and corresponding *entity_uri*.

```
herd = HERD()
(1) herd.add_ref(file=..., # The file containing the objects
                 container=..., # The object containing the data or attribute
                 attribute=..., # The attribute being linked
                 field=..., # Only used for compound data
                 key=..., # The value of the target data or attribute
                 entity_id=..., # The id from the resource for the entity
                 entity_uri=...) # The id from the resource for the entity

(2) herd.add_ref_termset(file=...,
                         container=...,
                         attribute=...,
                         termset=...)
```

**Figure 5** Overview of all the parameters in the *add_ref* and *add_ref_termset* methods the user would have to manage when populating an instance of *HERD.* Internally, both methods are able to resolve the file from the container if it has been added to the file, further reducing parameters.

To minimize the number of parameters users have to specify manually when creating references, we integrate the *TermSet* class with *HERD*. A term set stores a relationship between a user defined term, i.e., a *HERD key*, and the ontology defined term ID, i.e., a *HERD entity_id. TermSet* is also able to retrieve the *entity_uri* through relating the ID to the prefix. As **Figure 5** (step 2) shows, using this approach significantly simplifies the set of parameters a user needs to set manually. In addition, the *add_ref_termset* method also supports the *key* field to be a single term or a list of terms or a whole dataset. The method in turn then iterates over all the values to create the metadata linkages in a bulk-fashion. Regardless of the number of keys provided, each key value is also validated according to the values in the *TermSet*. The *TermSet* acts as a controlled set of permissible values when populating the *HERD* structure; invalid terms are returned for the user to resolve. This method introduces the first step in automation and validation.

### 2.2.2 How to Annotate New Data and Metadata

Beyond contextualizing existing datasets, we also would like to ensure that every dataset created going forward can have these rich linkages by easily integrating a system of tools into the existing user workflow. In addition, we want to ensure that these datasets are validated according to expected experimental outcomes and definitions. To address this need we created the *TermSetWrapper* class in HDMF (see **Figure 6**). By using this wrapper, users can validate values from datasets and attributes according to a *TermSet*, while also acting as a flag to automatically populate and write an instance of *HERD* when the user writes the file. The wrapper does not interfere with I/O processes to read or write a NWB file, nor does it require any changes to data type validation for any existing or future NWB or HDMF objects. Validation is conducted before the actual NWB or HDMF object is instantiated, by iteratively checking over the values in the wrapped object according to the permissible terms in the *TermSet*. All values need to pass this validation for the instance to be created; all invalid terms are reported to the user for manual adjustments. New data is also validated when appended to datasets that have been wrapped. The wrapper class is fully integrated with HDMF data types, requiring no changes in how developers or users create and interact with these objects' internal methods.

```
# Create file as usual
nwbfile = NWBFile(...)
# Define TermSet
termset = TermSet(term_schema_path="...")
# Define Wrapped Attribute for Subject
wrapped_species = TermSetWrapper(value="Mus musculus", termset=termset)
subject = Subject(
    subject_id="001",
    age="P90D",
    description="mouse 5",
    species=wrapped_species
    sex="M",
)
nwbfile.subject = subject
```

**Figure 6** An example workflow for using *TermSetWrapper* for validating the *species* attribute in *Subject.*

Within an NWB file, multiple datasets and attributes can be wrapped with the *TermSetWrapper*, each using a different *TermSet*. HDMF data I/O classes (e.g., *NWBHDF5IO* in PyNWB) support writing the data file and *HERD* simultaneously. This functionality is implemented in the *HDMFIO* base I/O class in HDMF, such that HERD can be used with any HDMF backend, e.g., Zarr backend as part of the hdmf_zarr library. On write, users have a choice to either provide a *HERD* instance to the write method or use a new *HERD* instance generated as default. When the user calls the write method from the I/O, the internal method *add_ref_container* is called automatically to search through the file hierarchy for any instance of the *TermSetWrapper*. Each wrapped object is then resolved to provide the *add_ref* method the necessary *entity_id* and entity_uri from the *TermSet* associated. Using this approach, the user never has to manually add any references or provide parameters to add_ref for each target value after they create their file, further automating the external resources metadata workflow.

### 2.2.3 How to Use Terminologies towards Experimental Design

When designing an experiment, there will be fields in NWB neurodata types that the user would want validated according to a *TermSet*. To avoid manually wrapping each field for every instance across multiple sessions, i.e., multiple NWB files, we developed the *TypeConfigurator* class in HDMF. The *TypeConfigurator* takes YAML configuration files as input that defines associations between a field from a neurodata type in the NWB schema to a *TermSet* (see **Figure 7**). This configuration file also stores the namespace for the neurodata types, allowing users to define configurations across multiple namespaces, enabling configuration of term sets also for arbitrary extensions to the NWB schema. The configuration file itself represents the validation parameters a researcher is setting for the various data modalities within the experiment.

```yaml
namespaces:
  core:
    version: 2.7.0
    data_types:
      Subject:
        species:
          termset: nwb_subject_termset.yaml
      NWBFile:
        experimenter:
          termset: experimenter_termset.yaml
      ElectrodeGroup:
        location:
          termset: location_termset.yaml
      Device:
        manufacturer:
          termset: manufacturer_termset.yaml
```

**Figure 7** An example configuration file outlining unique TermSet instances for each field.

The user does not directly use the *TypeConfigurator* class, but in practice only creates and loads the configuration file with the *load_type_config* method (or unloads a configuration with the *unload_type_config* method). Once a configuration has been loaded, the user proceeds as usual with populating the NWB file with data; specifically without having to manually create and specify the *TermSetWrapper*. Hence, the only change required to incorporate type configuration into existing scripts is to load the configuration before the new NWB file is being instantiated. With a configuration loaded, every field that is associated with a *TermSet* in the configuration file will automatically be wrapped with a corresponding *TermSetWrapper* when creating a new data object. This provides the benefits and functionalities of wrapping with *TermSetWrapper* without having to manually wrap the fields themselves. This approach in turn also introduces a greater strictness with regard to enforcement of terminology. Since every instance of a data type defined in the configuration file is being validated to a set of permissible values, this means that a user can only create files that contain valid terms.

### 2.2.4 Using Terminologies in the Design of Data Standards

When defining community data standards or creating an extension to a standard for a new data modality, it is useful to enforce controlled terms whenever possible to further enhance rigor of the data. One approach is to include type configurations as described in the previous section along with the data schema to suggest the usage of particular term sets. To support more strict enforcement of controlled terminologies, we plan to further enhance the HDMF data modeling language to enable the use of term sets directly in data schema.

# 3. Results

## 3.1 Annotating Existing files

As described prior, *HERD* is designed to support adding linkages retroactively to data and attributes within existing files without having to rewrite or modify the original data. With NWB seeing international adoption and with close collaboration with the DANDI Archive, there are already a considerable number of datasets that are publicly available to the broader community. However, these dataset do not necessarily share the same standard terminology to represent important experimental parameters.

In **Figure 8**, we have three NWB files stored in the DANDI archive: sub-Rat203_ecephys.nwb, sub-EE_ses-EE-042_ecephys.nwb, and sub-BH243.nwb [16]. Each file has a *Subject,* containing the field *species*. This is a free-form text attribute, giving researchers the flexibility to denote the species according to any internal nomenclature. Even though it gives researchers flexibility, lab-specific terminology does not always translate to clear metadata. In these files, *Subject species* is Rattus norvegicus; however, each file referred to the same species under a different denomination: \`rat\`, \`rattus norvegicus\`, and \`rattus norvegicus domestica\`. To link the species to an ontological resource, e.g., the NCBI Taxonomy, researchers can either i) use add_ref to directly populate an instance of *HERD* for each *Subject species* or ii) generate a reusable *TermSet* to minimize the number of input parameters. This represents a larger, more general issue where researchers use different terms to mean the same thing, creating confusion when data is shared externally. By linking the different terms to the same unique entity in an ontology, disambiguates the differences between terms and provides a clear definition of their meaning.

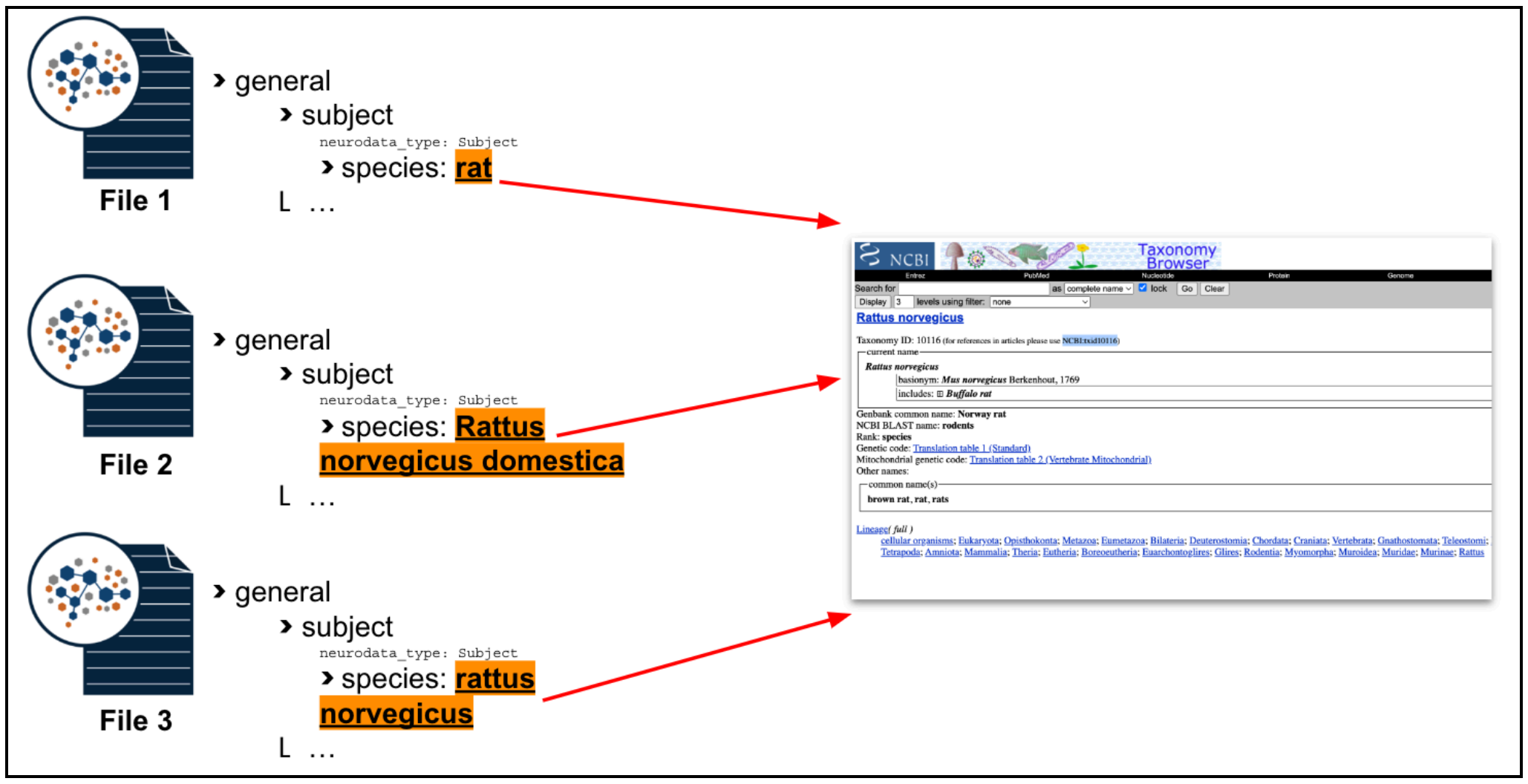

**Figure 8** Each of the three NWB files shown here contains different terms for the same species. *HERD* resolves any ambiguity for objects across different files by linking the different terms that are being used to the same ontological entity that uniquely identifies and rigorously describes the term.

## 3.2 Adding Linkages To New Data

The NWB standard has a core set of required and optional fields that users set when creating a new file. For example, it is common to store relevant researchers in the *experimenter* field. Laboratories can have an internal *TermSet* that stores the terms, i.e., the names of the researcher, and the meaning of each term, i.e., the corresponding ORCID identifier for each researcher. The process is identical when wrapping the *Subject species* field, as seen prior in **Figure 7**.

*HERD* also supports more sophisticated use cases, such as providing metadata linkages for electrophysiology. To store extracellular electrophysiology data within NWB, users must first create an electrodes table describing the electrodes used in the experiment. The electrodes table references a group of electrodes that belong to an *ElectrodeGroup*, which stores device information via the *Device* object. In this scenario, the *Device* used in the experiment can be linked to the device manufacturer for information regarding the device specifications, while the *location* field, representing the insert location of the probe, can be linked to a variety of brain atlases (see **Figure 9**).

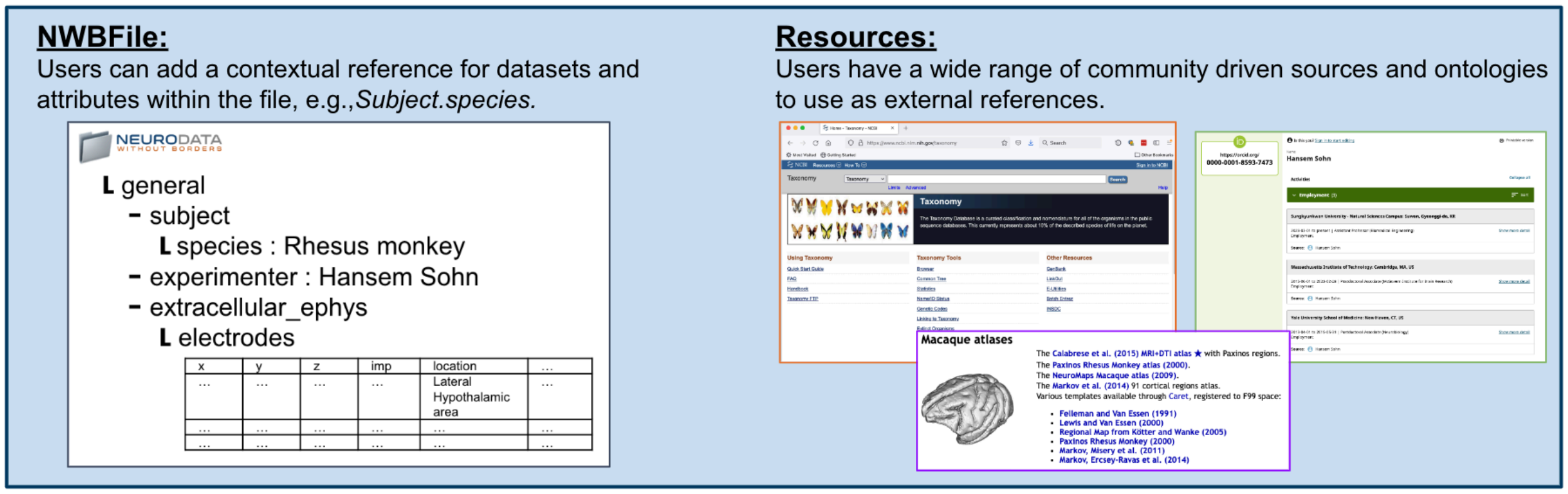

**Figure 9** The NWB ecosystem supports a variety of domains: extracellular electrophysiology, calcium imaging, intracellular electrophysiology, behavioral data, etc. *HERD* is able to connect any dataset/attribute for these domains to a community resource.

Using data from the DANDI sets from the previous section, we first create a *Device*, wrapping the manufacturer attribute with a *TermSetWrapper* (see **Figure 10**). Within the NWBFile, we create an *ElectrodeGroup*, defining both the device used and the brain area location, which are also wrapped. Alternatively, researchers can create a configuration YAML file that defines both the manufacturer field in *Device* and the *location* field in *ElectrodeGroup*, loading the configuration file prior to populating the file.

```
nwbfile = NWBFile(...)

device_termset = Termset(term_schema_path=...)
location_termset = Termset(term_schema_path=...)

device = nwbfile.create_device(name="electrode_probe_1",
                               description="Linear probe with 24 recording channels",
                               manufacturer=TermSetWrapper(value="Plexon Inc.",
                                                           termset=device_termset)

group = nwbfile.create_electrode_group(name="electrode_group_1”,
                                       description="Electrodes on a neural probe”
                                       device=device,
                                       location=TermSetWrapper(value="Dorsomedial frontal cortex",
                                                               termset=location_termset)
```

**Figure 9** For a complete description of the parameters for the *NWBFile* using this real dataset, refer to **Appendix A.3**.

# 4. Discussion

Creating a system to link, manage, and query contextual metadata that is flexible enough to encompass the multitude of neuroscience applications is no small challenge. The system needs to support not only the direct use-case of linking data to an external resource, but also provide metadata support to experimental attributes and parameters. The difficulty is compounded further by the sheer number of existing data in circulation, requiring the system to adopt methods that won’t require intensive alterations and I/O processing. We present *HERD* as a first step in accomplishing this task and as a strong base to continually evolve from and adapt to the community needs. At the core, our approach defines a standard data model for describing: 1) linkages between terms in files to external resources (*HERD*), and 2) curated collection of controlled terms that are defined according to an ontology (*TermSet*). Surrounding these standards are user-facing tools that aid researchers in using these methods to easily validate data and to formally define terms to disambiguate their experiments, i.e., *TermSetWrapper* and *TypeConfigurator*.

## 4.1 Future Work

The future will involve expanding the *HERD* suite of tools to more easily take advantage of *HERD* and *TermSet*, formalizing a community standard on how and where to store *HERD* files, and integrating the use of *HERD* into NWB extensions.

The next expansion in tools is to allow users to define the *TermSet* for the attribute or dataset directly in the NWB schema as mentioned in **Sec. 2.2.4**. This rigorous definition is similar to the *TypeConfigurator* in that new instances of neurodata type will have the defined fields validated and wrapped with the *TermSetWrapper*. Unlike the *TypeConfigurator*, however, users will not be able to toggle this validation, but definitions of terms defined in the data schema will always be enforced.

As mentioned prior, *HERD* is not written inside the NWB file*,* but is stored separately as part of the I/O write process for the file itself. The DANDI Archive currently houses hundreds of datasets that use the NWB format. In the near future, we want users to be able to store their exported *HERD* tables on the archive. We plan to work with DANDI to help define the policies and format requirements necessary to support sharing of *HERD* files via the archive. We also intend for the user-created termsets to be shareable for reuse across labs, requiring future development on where to store these sets.

## 6. Data Availability

All datasets used are available in the DANDI Archive [16a, 16b, 16c].

## 7. Code Availability

All NWB and HDMF codes and schema are available open source via the Neurodata Without Borders GitHub organization at https://github.com/NeurodataWithoutBorders and the HDMF GitHub organization at https://github.com/hdmf-dev, respectively. All codes are available under a permissive BSD license and all core software can be easily installed via common package managers, such as conda and pip. All documentation and training resources are available online via the https://www.nwb.org/ website. This manuscript is based on the following schema and software versions: i) NWB schema [11], ii) hdmf-common-schema [12] iii) HDMF [1], iv) PyNWB [2], v) MatNWB [13].

## Author Contributions

Conceptualization: OR, MA, AT, RL
Methodology: OR, MA, RL
Software: MA, RL, AT, OR, BD
Writing - Original Draft: MA
Writing - Review & Editing: MA, OR, RL, KB
Visualization: MA
Supervision:
Project Administration:
Funding Acquisition: OR, BD

## Competing Interests

The author(s) declare no competing interests.

## Acknowledgements

Research reported in this publication was supported by the National Institute Of Neurological Disorders And Stroke of the National Institutes of Health under Award Number U24NS120057 to O. Rübel and B. Dichter. Additional support for this research was provided by the Kavli foundation and by the Simons Foundation for the Global Brain grant 521921 to L. Frank. The content is solely the responsibility of the authors and does not necessarily represent the official views of the National Institutes of Health, the Kavli Foundation, or the Simons Foundation.

We thank all members of past and current NWB community working groups (see https://www.nwb.org/working-groups/) as well as all participants of the NWB hackathon series and the NWB user and developer community for their feedback and contributions. We thank the current and former members of the NWB Executive Board: Kristofer Bouchard, Bing Brunton, Elizabeth Buffalo, Anne Churchland, Loren Frank, Satrajit Ghosh, Adam Kepecs, Mala Murthy, Huib Mansvelder, Ueli Rutishauser, Lydia Ng, Christof Koch, Friedrich Sommer, Karel Svoboda, Markus Meister, and Katrin Amunts.

## Disclaimer

# Appendix

## A.1 TermSet

```yaml
id: termset/species_example
name: Species
version: 0.0.1
prefixes:
  NCBI_TAXON: https://www.ncbi.nlm.nih.gov/Taxonomy/Browser/wwwtax.cgi?mode=Info&id=
imports:
  - linkml:types
default_range: string

enums:
  Species:
    permissible_values:
      Homo sapiens:
        description: the species is human
        meaning: NCBI_TAXON:9606
      Mus musculus:
        description: the species is a house mouse
        meaning: NCBI_TAXON:10090
      Ursus arctos horribilis:
        description: the species is a grizzly bear
        meaning: NCBI_TAXON:116960
      Myrmecophaga tridactyla:
        description: the species is an anteater
        meaning: NCBI_TAXON:71006
```

## A.2 DynamicEnumerations

### 2.1 Schema to Dynamically populate enumerations

```yaml
id: https://w3id.org/linkml/examples/nwb_dynamic_enums
name: Cell Ontology
description: This schema is used to dyamically generate a new schema.

prefixes:
  linkml: https://w3id.org/linkml/
  CL: http://purl.obolibrary.org/obo/CL_

imports:
  - linkml:types

default_range: string

# ========================= #
#          ENUMS            #
# ========================= #
enums:
  NeuronTypeEnum:
    reachable_from:
      source_ontology: obo:cl
      source_nodes:
        - CL:0000540  ## neuron
      include_self: false
      relationship_types:
        - rdfs:subClassOf
```

2.2 Generated Schema with Enumerations

```yaml
id: https://w3id.org/linkml/examples/nwb_dynamic_enums
name: Cell Ontology
description: This schema is the generated schema.

prefixes:
  linkml: https://w3id.org/linkml/
  CL: http://purl.obolibrary.org/obo/CL_

imports:
- linkml:types

default_range: string

# ======================== #
#          ENUMS           #
# ======================== #
enums:
  NeuronTypeEnum:
    reachable_from:
      source_ontology: obo:cl
      source_nodes:
      - CL:0000540    ## neuron
      include_self: false
      relationship_types:
      - rdfs:subClassOf
    permissible_values:
      CL:0000705:
        text: CL:0000705
        meaning: CL:0000705
        title: R6 photoreceptor cell
      CL:4023108:
        text: CL:4023108
        description: A magnocellular neurosecretory cell that is capable of producing
          and secreting oxytocin.
        meaning: CL:4023108
        title: oxytocin-secreting magnocellular cell
      CL:0004240:
        text: CL:0004240
        description: An amacrine cell with a wide dendritic field, dendrites in S1,
          and post-synaptic terminals in S1.
        meaning: CL:0004240
        title: WF1 amacrine cell
```

## A.3 Complete NWBFile fields with TermSetWrapper

```
start_time =datetime(2016, 12, 11, 0, 0, 0, tzinfo=pytz.timezone('US/Eastern')).strftime('%Y-%m-%d
%H:%M:%S%z')
nwbfile = NWBFile(
    session_description="Data from monkey Haydn performing ready-set-go time interval reproduction
task...",
    identifier="8d26aa92-3929-11ec-8077-43176b153428",
    session_start_time=start_time
    session_id="20161211",
    experimenter='Hansem Sohn'
    lab="Jazayeri",
    institution="Massachusetts Institute of Technology",
    experiment_description="Cognitive timing task in which subject attempts to reproduce interval
between two cues",

)

device_termset = Termset(term_schema_path=...)
location_termset = Termset(term_schema_path=...)

device = nwbfile.create_device(name="electrode_probe_1",
                               description="Linear probe with 24 recording channels",
                               manufacturer=TermSetWrapper(value="Plexon Inc.",
                                                           termset=device_termset)

group = nwbfile.create_electrode_group(name="electrode_group_1",
                                       description="Electrodes on a neural probe"
                                       device=device,
                                       location=TermSetWrapper(value="Dorsomedial frontal cortex",
                                                               termset=location_termset)
```